\documentclass[conference]{IEEEtran}
\IEEEoverridecommandlockouts
\usepackage{cite}
\usepackage{booktabs}
\usepackage{amsmath,amssymb,amsfonts}
\usepackage{algorithmic}
\usepackage{graphicx}
\usepackage{textcomp}
\usepackage{graphicx} 
\usepackage{xcolor}
\usepackage{caption}
\def\BibTeX{{\rm B\kern-.05em{\sc i\kern-.025em b}\kern-.08em
    T\kern-.1667em\lower.7ex\hbox{E}\kern-.125emX}}
\begin{document}

\title{Subject Disentanglement Neural Network for Speech Envelope Reconstruction from EEG
}

\author{\IEEEauthorblockN{1\textsuperscript{st} Li Zhang, 2\textsuperscript{nd} Jiyao Liu }
 
\IEEEauthorblockA{\textit{Audio, Speech and Language Processing Group (ASLP), School of Computer Science} \\
\textit{Northwestern Polytechnical University (NPU), Xi’an, China} 
}

}

\maketitle

\begin{abstract}
Reconstructing speech envelopes from EEG signals is essential for exploring neural mechanisms underlying speech perception. Yet, EEG variability across subjects and physiological artifacts complicate accurate reconstruction. To address this problem, we introduce Subject Disentangling Neural Network (SDN-Net), which disentangles subject identity information from reconstructed speech envelopes to enhance cross-subject reconstruction accuracy. SDN-Net integrates three key components: MLA-Codec, MPN-MI, and CTA-MTDNN. The MLA-Codec, a fully convolutional neural network, decodes EEG signals into speech envelopes. The CTA-MTDNN module, a multi-scale time-delay neural network with channel and temporal attention, extracts subject identity features from EEG signals. Lastly, the MPN-MI module, a mutual information estimator with a multi-layer perceptron, supervises the removal of subject identity information from the reconstructed speech envelope. Experiments on the Auditory EEG Decoding Dataset demonstrate that SDN-Net achieves superior performance in inner- and cross-subject speech envelope reconstruction compared to recent state-of-the-art methods.  
\end{abstract}


\section{Introduction}
Speech envelope reconstruction from EEG signals is a growing research area focused on decoding speech-related information from EEG data \cite{pasley2012reconstructing}. This involves reconstructing the temporal envelope of speech, which reflects amplitude modulations over time, from EEG signals \cite{thornton2022robust, zuk2021envelope}. Applications span fields such as healthcare, education, entertainment, and more broadly, speech recognition and enhancement \cite{easwar2023predicting, nieto2022thinking}. Speech envelope reconstruction is critical as it decodes neural activity associated with speech processing \cite{keshavarzi2022decoding}.

Despite advancements using techniques like multivariate temporal response functions and neural coding in the auditory cortex \cite{mesgarani2008phoneme, lalor2006vespa, ding2012neural}, reconstructing speech envelopes from EEG remains challenging due to high inter-subject variability and EEG’s subject dependency, influenced by mental states, electrode impedance, and individual anatomical differences \cite{samek2013transferring, zhao2021plug}.
To address the challenge of subject-specific variability in EEG-based speech envelope reconstruction, we propose SDN-Net, which disentangles subject-specific information from EEG signals for subject-independent reconstruction.

The contributions of this paper are as follows:
\begin{itemize}
    \item We introduce a multi-level aggregation EEG codec (MLA-Codec) for accurate speech envelope decoding from EEG signals.
    \item We design a multi-scale time-delay neural network with channel and temporal attention (CTA-MTDNN) for effective subject classification, extracting subject identity from EEG.
    \item We integrate a mutual information estimator with a multi-layer perceptron (MPN-MI) to combine envelope reconstruction and subject classification, mitigating cross-subject variability in the reconstruction process.
\end{itemize}

\begin{figure*}[h]
  \centering
  \includegraphics[width=\linewidth]{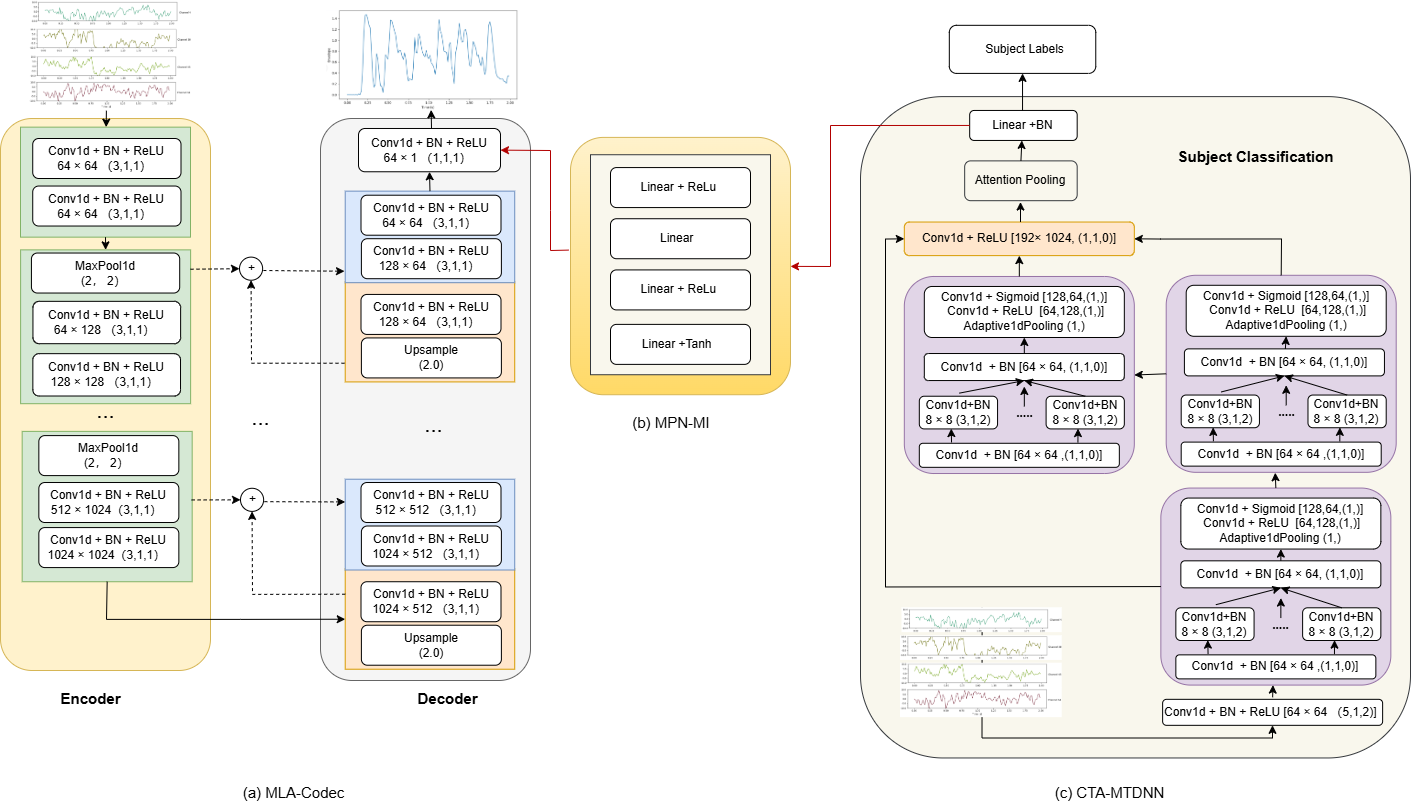}
  \caption{The overview of SDN-Net. (a) is a multi-level aggregation EEG codec (MLA-
Codec) for decoding speech envelopes. (b) is a mutual information (MI) estimator with a multi-layer perceptron network (MPN-MI). (c) is a multi-scale time-delay neural network with channel and temporal attention (CTA-MTDNN)  for subject classification. \textit{Note: The signals of four channels (Actually, there are 64-channel EEG signals) as the input is simply drawn to save space.}}
    \label{fig:overview}
\end{figure*}
\section{Methodology}
SDN-Net is an end-to-end neural network designed to reconstruct speech envelopes from EEG signals, improving cross-subject accuracy through mutual information disentanglement. As shown in Fig. \ref{fig:overview}, SDN-Net consists of three main modules: MLA-Codec for EEG encoding and decoding, MPN-MI for mutual information estimation, and CTA-MTDNN for subject classification with multi-scale time-delay, channel, and temporal attention.

The input to SDN-Net is 64-channel EEG data processed according to \cite{accou2023decoding,new7478117}, shown with four channels for simplicity in Fig. \ref{fig:overview}. Let $X = \{X_1, \dots, X_N\}$ represent EEG samples, $R = \{R_1, \dots, R_N\}$ the corresponding speech envelopes, and $Y = \{Y_1, \dots, Y_N\}$ the subject labels. The MLA-Codec for speech envelope reconstruction is defined as $G_\theta(X_i)$, CTA-MTDNN for subject classification as $C_\delta(X_i)$, and the MPN-MI mutual information estimator as $M_\alpha(G_\theta(X_i)|C_\delta(X_i))$, with learnable parameters $\theta$, $\delta$, and $\alpha$.

\subsection{Envelope Reconstruction}
As shown in Fig. \ref{fig:overview} (a), the MLA-Codec comprises an EEG encoder and decoder. The encoder consists of five convolution and max-pooling modules (highlighted in green), denoted as $G_{\theta}^{e^j}(X_j)$ for each module $j = 1,2,3,4,5$. The encoder's output can be expressed as:
\[
E(X_i) = G_{\theta}^{e^5}(G_{\theta}^{e^4}(G_{\theta}^{e^3}(G_{\theta}^{e^2}(G_{\theta}^{e^1}(X_i)))))
\]
This encoding process generates five bottleneck representations that capture both high- and shallow-level features for reconstructing the speech envelope.

The decoder consists of upsampling and convolution layers (in yellow) and additional convolution layers $G_{\theta}^{d^{j'}}$ (in blue) for refinement, where $j' = 1,2,3,4$. The upsampling layers increase temporal resolution, and envelope-related information is refined and concatenated with encoder outputs via $\phi$, as follows:
\[
D_{j'}(X_i) = \phi (G_\theta^{d^{j'}}, G_\theta^{e^1})
\]
The final output is decoded by a convolution layer $G_\theta^{d^{5'}}$, producing the reconstructed speech envelope $R_i'$:
\[
R_i^{'} = G_\theta^{d^{5'}}(X_i)
\]
The reconstructed envelope $R_i'$ is compared with the ground truth $R_i$ using a Pearson correlation coefficient loss, $L_{\text{corr}}$, defined as:
\[
L_{\text {corr }}=1 - \frac{\sum_{i=1}^N\left(R_i - \overline{R}\right)\left( R'_i - \overline{R'}\right)}{\sqrt{\sum_{i=1}^N\left(R_i - \overline{R}\right)^2} \sqrt{\sum_{i=1}^N\left( R'_i - \overline{R'}\right)^2}}
\]
where $\overline{R}$ and $\overline{R'}$ are the means of the true and predicted speech envelopes, respectively, and $N$ is the sample count.
\subsection{Subject Classification}
In SDN-Net, subject classification relies on a multi-scale time-delay neural network with channel and temporal attention, implemented as an enhanced 1D Res2Net block with squeeze-and-excitation (1D-SE-Res2Net), illustrated in purple in Fig. \ref{fig:overview}(c). Each 1D-SE-Res2Net block is denoted as $C_\delta^j(X_i)$ for $j = 1, 2, 3$, and its multi-scale output is:
\[
C'(X_i) = \phi \left(C_\delta^1(X_i), C_\delta^2(C_\delta^1(X_i)), C_\delta^3(C_\delta^2(C_\delta^1(X_i)))\right)
\]
where $\phi$ represents concatenation. This multi-scale representation then passes through a convolutional layer $C_\delta^4$, resulting in:
\[
C_\delta^5(X_i) = C_\delta^4(C'(X_i))
\]
where $C_\delta^5(X_i) \in \mathbb{R}^{C \times T}$, with $C$ and $T$ denoting the channel count and frame count, respectively.

To capture subject-specific EEG features, an attention module (in yellow) applies self-attention across time and channels. Using parameters $W \in \mathbb{R}^{R \times C}$ and $b \in \mathbb{R}^{R \times 1}$, the representation is projected to a lower-dimensional space, reducing overfitting risk. A non-linearity $\delta(x)$ transforms the representation to channel-specific self-attention scores via a linear layer with weights $v_c^T \in \mathbb{R}^{R \times 1}$ and bias $k_c$:
\[
e_{t, c} = \boldsymbol{v}_c^T \delta\left(\boldsymbol{W} C_\delta^5(X_i) + \boldsymbol{b}\right) + k_c
\]
The score $e_{t, c}$ is normalized over all frames with a softmax function:
\[
\alpha_{t, c} = \frac{\exp \left(e_{t, c}\right)}{\sum_{\tau=1}^n \exp \left(e_{\tau, c}\right)}
\]
The self-attention score $\alpha_{t, c}$ represents frame importance for each channel. The weighted mean $\tilde{\mu}_c$ for channel $c$ is computed as:
\[
\tilde{\mu}_c = \sum_{t=1}^n \alpha_{t, c} C_\delta^5(X_i)_{t, c}
\]
The weighted standard deviation $\tilde{\sigma}_c$ is:
\[
\tilde{\sigma}_c = \sqrt{\sum_{t=1}^n \alpha_{t, c} C_\delta^5(X_i)_{t, c}^2 - \tilde{\mu}_c^2}
\]
The final output of the attention module is the concatenation of $\tilde{\mu}$ and $\tilde{\sigma}$:
\[
C_\delta^6(X_i) = \phi(\tilde{\mu}, \tilde{\sigma})
\]
where $\phi$ denotes concatenation. The subject classification loss, measured by cross-entropy, is:
\[
L_{\text{subject}} = -\left[Y \log C_\delta^6(X) + (1 - Y) \log (1 - C_\delta^6(X))\right]
\]
 
\subsection{Mutual Information Estimator for Disentangle Subject Identity}
To address subject-specific information in EEG-based speech envelope reconstruction, we propose SDN-Net. First, we pre-train the subject classification network and then fix its parameters to serve as a subject representation extractor, $C_\delta(X_i)$. Next, we apply the MI estimator network, $M_\alpha(G_\theta(X_i) | C_\delta(X_i))$, using the variational MI estimation method from \cite{cheng2020club}. The variational MI loss, $\mathrm{L}_{\mathrm{var}}$, is calculated as the difference between expected log-probabilities for subject and envelope representations, helping disentangle subject identity.

For each sample, we approximate $\mathrm{L}_{\mathrm{var}}$ by comparing subject-related probabilities across pairs of samples. The total objective function combines the reconstruction loss $L_{\text{corr}}$, estimated MI $\hat{L}_{\mathrm{var}}$, and a clipped MI penalty term to ensure non-negativity, optimized as:

\[
L_{\text{total}} = \min_{f(x)} \max_{M_\alpha} \left(L_{\text{corr}} + \hat{L}_{\mathrm{var}} + \lambda \cdot \max(0, L_{\mathrm{var}})\right)
\]

where $\lambda$ balances these losses. Following pre-training of the subject classifier, we iteratively train the MI estimator and reconstruction network, enhancing the MLA-Codec’s subject-independent reconstruction capability.
\section{Experimental Results and Analysis}
SDN-Net includes three main modules—MLA-Codec for speech envelope decoding, MPN-MI for mutual information estimation, and CTA-MTDNN for subject classification—each validated experimentally.

\subsection{Results of MLA-Codec }
\begin{figure}[th]
  \centering
\includegraphics[width=\linewidth]{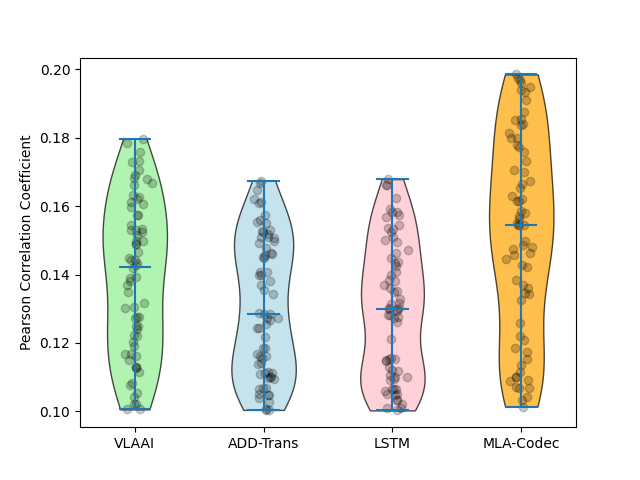}
  \caption{Mean of Pearson Correlation Coefficient of Different Models on Different Subjects.}
  \label{fig:res0}
\end{figure}

We compared the performance of our MLA-Codec model against state-of-the-art models, including VLAAI \cite{accou2023decoding}, ADD-transformer \cite{xu2022decoding}, and LSTM \cite{monesi2020lstm}, using Pearson correlation values between reconstructed and ground-truth speech envelopes. As shown in Fig. \ref{fig:res0}, MLA-Codec consistently outperformed the other models.

\begin{figure}[th]
  \centering
\includegraphics[width=\linewidth]{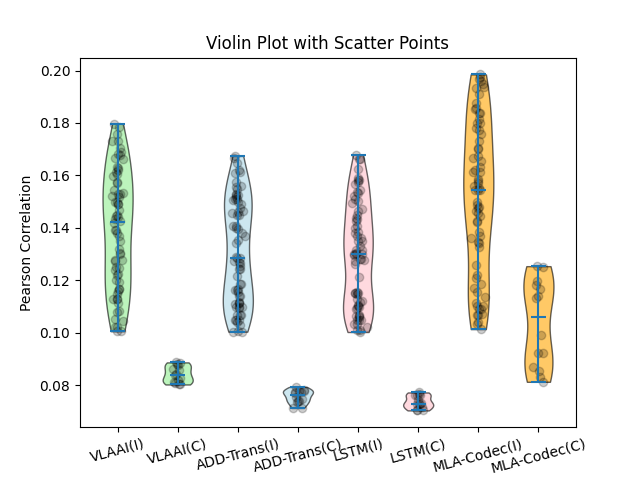}
  \caption{Mean of Pearson Correlation Coefficient of Cross-subject and inner-subject. (I) means inner-subject test results. and (C) means cross-subject test results.}
  \label{fig:res01}
\end{figure}

Additionally, we conducted an analysis of the inner-subject and cross-subject performance of various models, as presented in Fig. \ref{fig:res01}. The results demonstrate a notable difference in speech envelope reconstruction accuracy between inner-subject and cross-subject performance. Consequently, we propose to employ mutual information estimation techniques to disentangle subject representations to enhance the envelope reconstruction accuracy of cross-subject.

\subsection{Results of CTA-MTDNN}
In order to evaluate the performance of our proposed CTA-MTDNN module for subject classification, we compare it with two state-of-the-art methods: ES1D \
\cite{arnau2017es1d} and Multi-object \cite{moctezuma2020multi}. The results of the comparison are presented in Table \ref{tab:my-table}, which shows that the proposed CTA-MTDNN module outperforms the other two methods in terms of subject classification accuracy. This indicates the effectiveness of the proposed multi-scale time-delay neural network with channel and temporal attention mechanisms in extracting subject identity information from EEG signals.
\begin{table}[ht]
\centering
{\fontsize{7pt}{9.6pt}\selectfont 
\setlength\arrayrulewidth{0.5pt} 
\setlength{\heavyrulewidth}{0.5pt} 
\setlength{\aboverulesep}{0.0pt} 
\setlength{\belowrulesep}{0.0pt} 
\caption{The Accuracy of Subject Classification}
\label{tab:my-table}

\begin{tabular}{llll}
\hline
\multicolumn{1}{l|}{Model Name} & \multicolumn{1}{l|}{Accuracy (\%)} & \multicolumn{1}{l|}{Mean (\%)} & Std (\%) \\ \hline
ES1D \cite{arnau2017es1d}                          & 97.6                               & 95.3                           & 0.32     \\ \hline
Multi-object \cite{moctezuma2020multi}                  & 98.0                               & 96.1                           & 0.51     \\ \hline
CTA-MTDNN                       & 99.8                               & 97.2                           & 0.26     \\ \hline
\end{tabular}}
\end{table}

\subsection{Results of MPN-MI}
\begin{figure}[th]
  \centering
  \includegraphics[width=\linewidth]{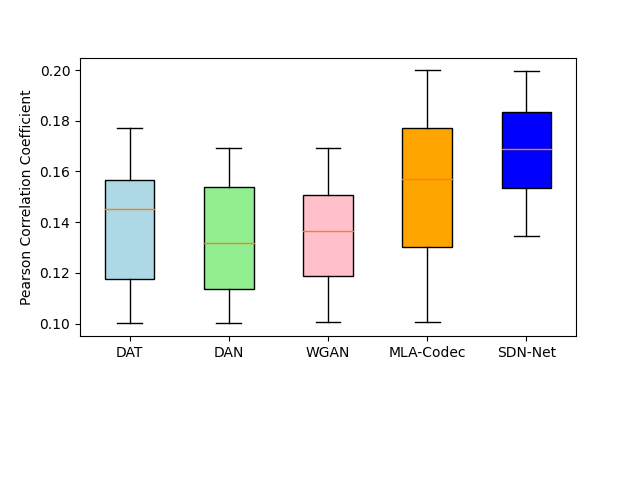}
  \caption{Mean of Pearson Correlation of Different Generalization Model for Cross-subject Challenge.}
  \label{fig:res08}
\end{figure}

To address the issue of reduced performance caused by inter-subject variability, we propose the MPN-MI module, which disentangles subject identity information from the reconstructed speech envelopes. By integrating the MLA-Codec, CTA-MTDNN, and MPN-MI modules, we obtain the complete SDN-Net framework. We evaluate the performance of SDN-Net against other model generalization methods, including DAT, DAN, and WGAN. The Pearson Correlation Coefficients of the different methods are presented in Fig. \ref{fig:res08}. According to the results in Fig. \ref{fig:res08}, the proposed SDN-Net method yields higher accuracy in reconstructing the speech envelope.

\section{Conclusion}

We propose a neural network architecture, SDN-Net, to tackle cross-subject speech envelope reconstruction. SDN-Net includes three main components: a multi-level aggregation EEG codec (MLA-Codec) for decoding speech envelopes, a mutual information estimator (MPN-MI) for supervising envelope reconstruction without subject identity, and a multi-scale time-delay neural network with channel and temporal attention (CTA-MTDNN) for subject classification. MLA-Codec encodes EEG signals into speech envelopes using a fully convolutional network, while CTA-MTDNN extracts subject identity through multi-scale learning and attention mechanisms. Experiments on the Auditory EEG Decoding Dataset show that SDN-Net surpasses recent methods in speech envelope reconstruction for both inner- and cross-subject test sets.
\bibliographystyle{unsrt}
\bibliography{sample-base}


\end{document}